-------------------------------------------------------------------------------
\documentstyle[twocolumn,aps]{revtex}
\begin{document}
\draft
\preprint{TIFR/TH/95-37}
\title{Active-Site Motion and Pattern Formation in Self-Organised
Interface Depinning}
\author{Supriya Krishnamurthy and Mustansir Barma}
\address{Theoretical Physics Group, Tata Institute of Fundamental Research,
Homi Bhabha Road, Bombay 400005, India \\
e-mail: skrish,barma@theory.tifr.res.in}
\maketitle
\begin{abstract}
We study a dynamically generated pattern in height gradients,
centered around the active growth site, in the steady state of 
a self-organised interface depinning model. The pattern has a
power-law tail and depends on interface slope.
An approximate integral equation relates the profile to local interface
readjustments and long-ranged jumps of the active site.
The pattern  results in a two-point correlation function
saturating to a finite value which depends on system size.
Pattern formation is generic to systems in which
the dynamics leads to correlated motion of the active site. 
\end{abstract}

A common feature of many self-organised systems 
is that their dynamics is extremal \cite{mm}: 
that is, activity is triggered at that spot where
a link is weakest, or a force the strongest, and not
statistically uniformly over the full system.
As a result, the active site performs an erratic motion
with nontrivial correlations in space and time \cite{pmb}. 
In this Letter, we demonstrate an intriguing and generic feature of
such systems: the existence 
of a well-defined spatial pattern around this
dynamical center of activity.

We study this phenomenon in a model of a self-organizing interface in a
random medium. In steady state, we find that there is a nonzero 
time-averaged spatial pattern of height differences, provided that 
it is referred always to the location of the active site at
that instant.
Figure 1 is a schematic representation of the interface profile,
time-averaged in a frame which keeps the active site at the origin.
Interfaces in which  $ r \rightarrow  -r$ symmetry is present (Fig. 1a), 
or absent (Fig. 1b), are shown, symmetry-breaking being induced
by tilting the interface \cite{tkd}. In the former case, the profile 
has a cusp
at the origin, while in the latter case the dominant feature is a
larger slope near the origin --- a local amplification of the
broken symmetry. The activity-centered pattern (ACP) is defined in
terms of height gradients,
\begin{equation}
\Psi (r) \equiv \frac{1}{2}[\langle \nabla h(r+R(t))\rangle - m]
\, , \label{eq:pattern}
\end{equation}
where $ R(t) $ is the position of the active site at time $t$,
$h(r^{\prime})$ denotes the
height at the site $r^{\prime}$, $\langle ... \rangle $ is a time
average and $m$ is the overall slope of the interface. 
>From numerical simulations, we find that $\Psi(r)$  has a long-ranged
(power law) tail with  exponents which differ in tilted and untilted
interfaces. We also find that the ACP leads to an unusual finite size
effect, namely the 
saturation of a two-point correlation function with increasing
separation. 
Finally, we present an approximate integral equation which relates
the ACP to the local dynamics of interface readjustment
and the long-ranged dynamics of active-site motion.
This provides an understanding of the form of the
induced pattern as a function of spatial correlations in the activity.

In our model the interface is taken to be a directed path on a
square lattice (Fig. 2), with tilted cylindrical boundary conditions 
\cite{mbjphysa} 
which ensure that the mean slope is preserved. To every bond
$k$ on the lattice is pre-assigned a fixed random number $f_k$ 
drawn from
the interval [0,1]. 
Motion of the interface is initiated at
that bond, on the forward perimeter of the interface, which carries 
the smallest random number.
In order that the length of the interface not change (an effect of surface 
tension), its
directed walk character 
is preserved by local re-adjustments;
if the chosen minimal bond has a positive (negative) slope, the
sequence of  links with negative
(positive) slope just below (on the left) also advance \cite{dlss},
as illustrated 
in Fig. 2.  This model resembles that introduced by 
Sneppen \cite{snep} but differs from it in that the length of the
interface is a strict constant of the motion and the $ f_{k}$'s 
are associated with bonds rather than sites.

The interface dynamics is equivalent to the dynamics 
of a system of hard-core particles on a ring, with a
positive-slope link of the interface represented by
a particle ($n_j =1$) and a negative-slope link by a hole ($n_j = 0$);
see Fig. 2. The difference in height of the interface between
sites $j_{1}$ and $j_{2}$ is given by 
$ h_{j_{2}} - h_{j_{1}} = \sum_{j=j_{1}}^{j_{2}} (2 n_{j} - 1)$.
Each site $j$ of the ring also carries the
random number $f_{j}$ assigned to the bond in front
of the corresponding spot on the interface.
In each time step, activity is
initiated at the site with the minimum $f_{j}$. If this site contains a 
particle (hole), it exchanges with the first hole (particle)
on the left (right). All sites hopped over, including the two which
exchange the particle and hole, are refreshed by assigning a new set of
$f_{j}$'s. The overall particle density
$\rho$ is strictly conserved, and determines the mean slope 
$m=2\rho -1$ of the interface. There is 
a nonzero current, which translates into a
forward-advancing interface.

>From time to time, the interface  aligns along  directed
spanning paths in a percolation problem,
as in the Sneppen model \cite{tl,tl1}. In our case, the
corresponding percolation problem is simpler, {\it viz.}
directed bond percolation (DP) \cite{kinzel} on the dual lattice, 
or equivalently,
diode-resistor percolation (DRP) \cite{dbp} on the original lattice. 
For a given
configuration $\{f_k\}$  and a trial threshold value
$f_o$, occupy all bonds $k^{\prime}$ with $f_{k^{\prime}}>f_o$ 
by diodes (one-way connections
pointing against the direction of advance of the interface, see Fig. 2),
and place resistors (two-way connections) on the remaining bonds. 
The  diode concentration is $p=1-f_o$. 
Infinite connected paths (stoppers) first form along the easy
direction of DP at  $p=p_{DP}
\simeq 0.6445$ on the square lattice.
An untilted interface ($m=0$) aligns along 
stoppers in the course of its motion \cite{tl,tl1}.
For $ m \neq 0$,
the corresponding threshold  $p_c(m) > p_{DP}$
is that value of $p$ for which the edge of the DP cone \cite{kinzel}
has slope $m$. As it evolves, a tilted interface of slope $m$ aligns along
stoppers of that slope \cite{amar}.

The pattern defined in (1) is associated with a
spatial profile of the density in the particle-hole 
model, as the height gradient maps on to the density.
The defining equation (1) now reads
$
\Psi (r) = \langle n(r+R(t))\rangle - \rho  
$.
We studied $\Psi (r)$ by Monte
Carlo simulation. 
In the untilted case, $ \Psi (r)$ is an odd
function (Fig. 3a) decaying asymptotically
as a power law $ |r| ^{-\theta}$  with
$ \theta = 0.90 \pm 0.03 $.  
In the tilted case
(Fig. 3b), as there is no $r \rightarrow -r$ symmetry, $\Psi(r)$ does
not have a definite parity. It is useful to separately analyse
$\Psi_{\pm}(r) \equiv (\Psi (r) \pm \Psi (-r))/2 $. The
odd part decays as $ \Psi_{-} (r) \sim  |r|^{- \theta_{-}} $ with $
\theta_{-} = 1.04 \pm 0.05 $. The even part $ \Psi_{+} (r) \approx
-b(L) + a ~|r|^{-\theta^{+}} $  where $ \theta_{+} = 0.46 \pm 0.05 $ and
$b(L) \rightarrow 0 $ as the lattice size $L \rightarrow  \infty$.

The density profiles of Fig. 3 correspond to the height patterns shown
in Fig. 1. 
In the untilted case, on average, the active
site is located at the peak (Fig. 1a)  where
$f_k$'s which have not been sampled earlier
are most likely to occur.
In the tilted case there is also a 
bootstrap effect at work.
If $ \rho \geq 0.5$, the active
site is more likely to contain a particle.
Thus regions to the left 
are more often refreshed, making
the active site likely to move in this direction
and find itself amidst a  particle cluster. This leads to densities 
higher than $\rho$ on both sides of the active site (Fig. 3b)
for $ \rho$ sufficiently different from $0.5$.

More quantitatively, we may write an integral equation for $\Psi(r)$
for untilted and tilted interfaces.
If the pattern is centered at $R(t)$ at time $t$, the dynamics 
causes two changes at the next instant:
(i) A short-ranged readjustment of the interface  changes the profile near
$R(t)$. As a result, 
the local density to the right of the active site becomes less than 
the density to its left. The average density
change at site $R(t)+r$ defines the density-increment function $ \Phi (r) $.
(ii) The active site jumps a distance $l \equiv R(t+1) - R(t) $; 
there is a substantial
probability $P(l)$ for large jumps. Since the ACP is centered at the
active site,
the result of (i) followed by (ii) is that the average profile
reproduces itself, except that it is centered at the shifted site 
$ R(t) + l $.
Both effects are incorporated into the  integral equation 
\begin{equation}
 \Psi (r) = \int_{-\infty} ^ {+\infty}  [ \Psi (r-l) + 
\Phi (r-l)] P(l) dl  \, . \label{eq:ACP}
\end{equation}
$P(l)$ decays as a power for large $l$:
$P(l) \sim |l|^{-\pi}$ (Fig. 4). In the
untilted case $P(l) = P(-l) $ holds because of 
$ r \rightarrow -r $  symmetry. We find
$ \pi = 2.25 \pm 0.05$ which compares well with earlier determined
values of $\pi$ for the Sneppen model \cite{snjn,tl1,pmb}. In the tilted 
case ($ \rho \neq 1/2$),
$ P(l)$ is not a symmetric function (Fig. 4) and it is convenient to
separately analyse the even and odd parts $ P_{\pm} \equiv 
(P(l) \pm P(-l))/2 $. We find that the even part, $ P_{+} (l) $, decays
asymptotically as $ P_{+} (l) \sim |l|^{ - \pi_ {+}}$ with $
\pi_{+} = 2.00 \pm  0.02$. The odd part $ P_{-} (l)$ changes sign 
(as implied by the crossing of the curves in Fig. 4) and asymptotically
follows $P_{-} (l) \sim |l|^{- \pi_{-}} $ with $ \pi_{-} =2.49 \pm
0.06$. We verified that the values of $ \pi_{+}$ and $\pi_{-}$ are the
same for various $ \rho \neq 1/2$. 

Equation (2) can be solved by Fourier transform. 
Defining $ \hat{\Psi 
}(q) \equiv  \int_{-\infty} ^ {+\infty}  e^ {2\pi i qr} \Psi(r) dr $
{\it etc} we find
\begin{equation}
\hat{\Psi} (q) = \frac {\hat{\Phi} (q) \hat{P} (q)} {1 - \hat{P} (q)}
\, .
\label{eq:FT}
\end{equation}
This equation indicates that the ACP arises as a
nonlocal response to the local density-increment function $
\Phi (r) $. Notice that the denominator vanishes as $q \rightarrow 0$.
We have solved (3) numerically, but it is more
instructive to examine the small-$q$ (large-$r$) behaviour.

In the untilted case, the $ r \rightarrow -r$ symmetry implies
that $ P(l)$ is even, while
$ \Phi(r) $ and the profile $ \Psi (r)$
are both odd. $ \Phi(r) $ is a short-ranged function with finite
first moment $ \phi_{1}$. Hence, to first order,
$ \hat{\Phi} (q) \approx i\phi_{1}
q$. The power-law
tails in $P(l)$ imply
that $ \hat{P} (q) \approx 1
- A |q|^{ \pi -1} $. Thus we find 
$ \Psi(r) \sim sgn(r) ~|r|^{ -(3- \pi)}$ {\it
i.e.} the profile is an odd function, with a power law tail. 

In the tilted case, the absence of $ r \rightarrow -r$
symmetry implies that none of $ P(l), ~\Phi(r) $ and $ \Psi(r)$ has a
definite parity. Since $ \Phi(r) $ is short-ranged, $ \hat{\Phi} (q)
\approx i\phi_{1} q + \phi_{2} q^{2}$ as $q \rightarrow 0$.
There is
no $ \phi_{0}$ term, as the elementary step of hopping a particle or
hole conserves particle number, implying $ \int \Phi (r) dr =0$. The $ q
\rightarrow 0 $ behaviour of $\hat{P} (q)$ is determined by the
asymptotic power law decays of the even and odd parts
$P_{\pm} (l)$ as $ |l| \rightarrow
\infty$. Thus we have $ \hat{P_{+}} (q)\approx 1 - A|q| ^{ \pi_{+}
-1}$. We might have expected $ \hat{P_{-}} (q) \approx  Bq +
C ~sgn (q) ~|q|^{ \pi_{-} -1} $, but in fact the mean velocity $\int
lP(l)dl $ of the active site vanishes \cite{foot} implying $B=0$. 
Thus the integral equation predicts that to leading order,
both $ \Psi_{+}(r)$ and $ \Psi_{-} (r) $ decay as powers $\sim $
$ |r|^{ -\theta_{\pm}}$, with $\theta_{+} + 2\pi_{+} -\pi_{-} = 3 $ and
$\theta_{-} + \pi_{+} = 3$. 
 The prediction $\pi_{+} +
\theta_{-} =3$
compares quite well with the numerical values $3.15$
for the untilted case and $3.04$ for  $\Psi_{-}(r)$  in the
tilted case. For $\Psi_{+}(r)$, however,
the numerically
determined value of $ \theta_{+} + 2\pi_{+} -\pi_{-} (\simeq 1.97)$
deviates more from the predicted value $3$. The reason behind the
discrepancy is that (2) implicitly ignores correlations
{\it e.g.} in successive jump lengths of the active site.

To check this point, we numerically studied a model with no
correlations in successive jump lengths. At every instant the jump
length of the active site was taken from a power-law distribution 
$P(l)$ chosen suitably so as to implement the zero velocity constraint.
Short-ranged adjustments were taken to follow
the same particle-hole hopping rules as the extremal model.
This model is
similar in spirit to the L\'{e}vy flight interface
model considered in \cite{snjn}. The results in this case 
were fully consistent
with the predictions  $\theta_{+} + 2\pi_{+} -\pi_{-} = \theta_{-} +
\pi_{+} = 3$. Further, we also studied cases in which $P(l)$ is not  a
power law. For short ranged $P(l)$,
the ACP has the form of a kink function, with a particle (hole) -
rich region on the left (right) of the active site.
In the 
limit of strictly infinite ranged $P(l)$, tantamount to normal
stochastic evolution \cite{dlss}, the ACP ceases to exist. 
In all cases we
compared our numerical results with those predicted by (2)
and found good agreement. 

The ACP has several interesting consequences.
For instance, we expect there to be a larger length of interface in a 
region of fixed size $x$
around the active site, than in a region opposite
it. Accordingly, we monitored mean squared fluctuations of the
height around the  instantaneous average,
in regions around and opposite the active site, and found a
pronounced difference (factor $ \simeq 2$, for tilted and untilted cases
with $x = 256$,$L=4096$).
This effect is smaller at stoppers, indicating that the ACP
itself is suppressed there.
The excess length of interface associated with
the ACP may provide a useful way to identify the active region in
experiment.     
 
Another consequence of the ACP is that being
a one-point correlation function, albeit an  unusual
one, it has a strong effect on  the
customary, space-time averaged two-point correlation function 
$ C(\Delta r) \equiv \{
{\langle n(r^{\prime})
n(r^{\prime}+\Delta r) \rangle} \}  -  \{ {\langle n(r^{\prime}) 
\rangle} {\langle n(r^{\prime}+\Delta r) 
\rangle} \} $. Here $r^{\prime}$ is a fixed site on the lattice,
$ \langle...\rangle $ stands for a time average 
and \{...\} stands for an average over  all sites $r^{\prime}$. 
Numerical results for $ C(\Delta r)$  (Fig. 5)  show that it 
saturates at a value $ C^{sat}$
which decreases with increasing size $L$
\cite{foot1}. 
This unusual behaviour can be understood in terms of the ACP.
Consider the correlation function $ \Gamma (r,\Delta r) = \langle
\delta n (r + R(t))\delta n( r+ \Delta r + R(t)) \rangle$
where $ R(t)$ is the location of the active site , $r$ is the
distance from the active site and $\delta n
(r +R(t)) \equiv n(r +R(t)) -\rho -\Psi(r)$
is the fluctuation
around the average ACP. A reasonable expectation is that the 
$\delta n$'s are independent for large separations
$\Delta r$, {\it i.e.}
$ \Gamma ( r,\Delta r) \rightarrow 0 $ as $ \Delta r \rightarrow
\infty$. On averaging over $r$ (which has the same effect as the
average over space-fixed sites $r^{\prime}$), we see that 
$ \{ {\langle n(r)n(r+\Delta r) \rangle} \} -
\{ {(\rho + \Psi (r))(\rho+ \Psi(r+ \Delta r))} \}
$ approaches zero as $\Delta r \rightarrow \infty$. This
predicts the saturation value $ C^{sat}$ to be
$\{ (\rho+ \Psi(r))(\rho+ \Psi(r +\Delta r)) \} - \rho^{2}
$. To test this, we subtracted this  from $C(\Delta r)$ 
and found that the saturation effect is in fact suppressed strongly 
(Fig. 5), supporting our interpretation \cite{foot2}. Customarily,
the large-separation saturation of $ C(\Delta r)$ is associated
with nonzero values of $ \langle n(r^{\prime}) \rangle - \rho$;
the unusual aspect here is that there is saturation even though
$ \langle n(r^{\prime}) \rangle = \rho$.

Though we have focussed on the ACP associated with a conserved
quantity like the density, a conservation law is not essential for the
occurrence of a nontrivial ACP.
For instance, on monitoring
the distribution of $ {f_{r}}$'s, with $r$ referred to the
active site, we find profiles with power-law tails 
in the interface depinning model with and without
tilt \cite{skmb}. We also find similar patterns
in the Bak-Sneppen model of evolution \cite{baksn}
and a modified asymmetric version of it \cite{skmb}. 
This indicates that pattern formation is a generic
feature  of systems evolving through  extremal dynamics.

In summary, we have explored a new aspect of self-organised
interface depinning, namely the existence of a spatial pattern
centered around the active site. We have shown that the pattern in
height gradients has power law tails which are sensitive to $r
\rightarrow  -r$ symmetry, broken by tilting the interface, and that
it explains the saturation of the two-point correlation function with
increasing separation. The integral equation (2) provides
an approximate description of the formation of the pattern.
We have also shown that power-law
patterns exist in other quantities, for instance the
distribution of random numbers around the minimum, in both the
interface and evolution models. Our studies suggest
that  systems which evolve through extremal dynamics are
likely to exhibit activity-centered patterns.

We thank  Gautam Menon and
Deepak Dhar for useful discussions, and  the referee for his
comments.

\centerline{\bf Figure Captions}

\begin{figure}[p]
\caption{Schematic view of averaged interface profile around the active
site A in (a) untilted (b) tilted interfaces. Tilt breaks $r
\rightarrow -r$ symmetry, an effect that is amplified near A.}
\end{figure}

\begin{figure}[p]
\caption{The bond model. The tilted interface $II^{\prime}$ 
advances along the extremal perimeter
bond $A$ and locally readjusts to align along the dashed line. At the
next instant, the activity moves from $A$ to $A^{\prime}$. The corresponding
configuration and local moves for the particle-hole model are also
shown. $ SS^{\prime}$ is a stopper, whose perimeter is fully occupied by
diodes.} 
\end{figure}

\begin{figure}
\caption{Density profiles in the (a) untilted ($\rho = 0.5$) 
and (b) tilted ($ \rho =0.75$) cases. 
Inset: The odd part of the profile in
tilted (open circles) and untilted (filled circles) cases.
We used $ L=16384 $ and averaged over $ 10^{8}$ configurations.}
\end{figure}

\begin{figure}
\caption{Monte Carlo results for the probability distribution 
of the jump of
the active site for three different densities, 
$\rho = 0.5$  (plus sign), 
$ \rho = 0.75$ (circles) and $\rho = 0.84375$
(triangles). If $\rho  \neq 0.5$, 
there is a left-right  asymmetry and the asymptotic slope differs from
that for $ \rho = 0.5$.
We used  L= 65536 and averaged over $3.10^{9}$
configurations.}
\end{figure} 

\begin{figure}
\caption{Density-density 
correlations for $L=4096$
(open circles) and for $L=16384$ (squares).
The saturation is reduced strongly (filled circles) on subtracting the
contribution of the
ACP.  We averaged over $ 10^{6}$ configurations.}
\end{figure}


\begin{thebibliography}{999}
\bibitem{foot} We numerically verified that the mean velocity vanishes.
This is related to the probability of first return of the activity to a site
decaying as a slow power of time.
\bibitem{foot2}
It is possible that another slightly different definition of the
pattern would eliminate the slight shoulder which remains after
subtraction (Fig. 5).
\end{thebibliography}
\end{document}